\documentclass[aps,prl,preprint,a4paper]{revtex4-1}

\usepackage{amsmath}
\usepackage{amssymb}
\usepackage{graphicx}
\usepackage{dcolumn}
\usepackage{bm}

\begin{document}


\title{The Hamiltonian Mechanics of Stochastic Acceleration} 



\author{J. W. Burby}
 \affiliation{Princeton Plasma Physics Laboratory, Princeton, New Jersey 08543, USA}
\author{A. I. Zhmoginov}
 \affiliation{Department of Physics, University of California, Berkeley, California 94720, USA}
\author{H. Qin}
 \affiliation{Princeton Plasma Physics Laboratory, Princeton, New Jersey 08543, USA}
 \affiliation{Dept. of Modern Physics, University of Science and Technology of China, Hefei, Anhui 230026, China}


\date{\today}

\begin{abstract}
We show how to find the physical Langevin equation describing the trajectories of particles undergoing collisionless stochastic acceleration. These stochastic differential equations retain not only one-, but two-particle statistics, and inherit the Hamiltonian nature of the underlying microscopic equations. This opens the door to using stochastic variational integrators to perform simulations of stochastic interactions such as Fermi acceleration. We illustrate the theory by applying it to two example problems.  



\end{abstract}


\maketitle 

\emph{Introduction}.\,--- The term ``stochastic acceleration" refers to the chaotic motion of particles subjected to a prescribed random force.  Such motion occurs in myriad contexts; the turbulent electromagnetic fields present in the interstellar medium and the RF wave fields found in magnetic fusion devices are just two examples. In the astrophysical context, it is thought to be partially responsible for the presence of cosmic rays in our solar system \cite{fermi49}. In the magnetic fusion context, it might explain the presence of certain high-energy tails observed in the National Spherical Torus Experiment when neutral beams are fired into RF-heated plasmas \cite{NSTX_HHFW}. 


Robust modeling of stochastic acceleration requires statistical approaches. The dominant approach is to employ the Fokker-Planck equation \cite{sturrock66,sturrock67,bar79,pet04,ham92} for the one-particle distribution function. However, when studying Richardson dispersion \cite{richardson,richpair}, and more generally any phenomenon governed by the two-particle distribution function \cite{two_point_exp}, the one-particle Fokker-Planck equation is insufficient. This is because spatial correlations in the random force field prevent the two-particle distribution function from factoring as a product of one-particle distribution functions. A superior statistical model when multi-particle statistics are in question would be a Langevin equation for particle trajectories. A wisely-chosen Langevin equation could capture the physics of the one- and two-particle distribution functions while providing an attractive means to perform Monte Carlo simulations of stochastic acceleration. Currently, there are no satisfactory methods for finding such a Langevin equation. 

The purpose of this Letter is to describe, for the first time, a systematic procedure for passing from a microscopic description of stochastic acceleration in terms of Hamiltonian equations of motion to the physically-correct Langevin equation for particle trajectories in the long-time limit. We will also show that, aside from reproducing the correct multi-particle statistics, this Langevin equation inherits the Hamiltonian structure of the microscopic dynamics. Specifically, we will show that the Langevin equation is a Hamiltonian stochastic differential equation (SDE) \cite{La08}. Thus, this work proves that symmetries of the macroscopic physical laws governing stochastic acceleration lead to conservation laws. 

We will focus our attention on stochastic acceleration problems similar to those studied in \cite{sturrock66,sturrock67,bar79,pet04}. These consist of a collection of non-interacting particles moving through a prescribed Hamiltonian force field. By assumption, the force will consist of a small-amplitude perturbation superimposed over a time-independent background. The perturbed force felt by a particle will be assumed to have a correlation time much shorter than any bounce time associated with the perturbation, zero mean, and temporally homogeneous statistics. These assumptions preclude treating Coulomb collisions because the polarization field produced by a particle cannot be modeled as a prescribed field; the polarization force depends on the history of a particle's orbit. They also preclude the treatment of strong turbulence \cite{dubois78}.  


\emph{The main idea.}\,--- Mathematically, this type of problem can be described as follows. Each particle moves through a $2n$-dimensional single-particle phase space $M$ according to a dynamical law given by a time-dependent vector field $X_t$; if $z_t\in M$ denotes the trajectory of a particle in $M$, then 
\begin{align}\label{micro}
\dot{z_t}=X_t(z_t).
\end{align} 
Because the only forces present are Hamiltonian, $X_t$ must be Hamiltonian in the sense that there is some Poisson bracket $\{\cdot,\cdot\}$ and some time-dependent Hamiltonian, $H_t$, such that  $\dot{z}^i=\{z^i,H_t\}$, where $z^i$ denotes an arbitrary coordinate system on $M$ \cite{littlejohn79}. By standard mathematical convention, this is written $X_t=X_{H_t}$  \cite{FoM}. The presumed form of the force then implies $H_t=H_0+\epsilon h_t$, where $\epsilon\ll1$, $H_0$ describes the mean time-independent background, and $h_t$ describes the small-amplitude random perturbation. Moreover, $X_{h_t}$ evaluated on a particle trajectory must have a correlation time $\tau_{\text{ac}}$ much shorter than some constant $\tau$, which, in turn, is much shorter than any bounce time associated with the perturbation $\tau_b$, $\tau_{\text{ac}}\ll\tau\ll\tau_b$.

Our goal in this Letter is to find the correct coarse-grained version of the microscopic equations of motion, $X_{H_t}$. Specifically, we seek a Langevin equation in the form 
\begin{align}\label{ansatz}
\delta z_t=X_0(z_t)\,\mathrm{d}t+\sum_{k\geq 1}X_k(z_t)\,\delta W^k_t
\end{align}
whose solutions correctly reproduce the late-time statistical behavior of solutions to the microscopic equations of motion.
Here $X_k$ are vector fields on $M$ that must be determined, $W^k$ are independent ordinary Wiener processes, and $\delta$ denotes the Stratonovich differential \cite{gardiner09} (sometimes also written $\circ \mathrm{d}$). We will identify the $X_k$ by demanding that Eq.\,(\ref{ansatz}) possess two properties: it must generate the Fokker-Planck equations for the one- and two-particle distribution functions, $f_t(z)$ and $g_t(z_1,z_2)$. The two-particle distribution function is defined such that the probability particle $1$ is in the region $U_1\subset M$ and particle 2 is in the region $U_2\subset M$ at time $t$ is given by $\int_{U_1}\int_{U_2}g_t\,\mathrm{d}z_1\,\mathrm{d}z_2$, where $\mathrm{d}z$ denotes the Liouville measure \cite{FoM}. Baxendale \cite{Bax84} has proven that a Langevin equation is uniquely determined by its one- and two- particle Fokker-Planck equations. Therefore, these conditions uniquely specify the Langevin equation we seek. In particular, the requirement that two-particle statistics be accurately reproduced is critical; Baxendale's work implies that constraining the Langevin equation only to be consistent with the one-particle Fokker-Planck equation would not identify it uniquely.

Physically, the reason that the two-particle Fokker-Planck equation contains more information than the one-particle Fokker-Planck equation can be understood as follows. After a short amount of time $\Delta t$, the displacement of a particle initially located at $z_1$ at time $t$ is given approximately by $\Delta t\, X_t(z_1)$. Similarly, the displacement of a particle initially located at $z_2$ is nearly $\Delta t\, X_t(z_2)$. Because the random force field generally has spatial correlations, $X_t(z_1)$ and $X_t(z_2)$ are not statistically independent. Thus, the probability distribution of $(z_1^\prime,z_2^\prime)$, where $z_i^\prime\approx z_i+ \Delta t \,X_t(z_i)$, will not be given by the product of the distribution of $z_1^\prime$ with that of $z_2^\prime$. This failure-to-factor precludes determining the two-particle distribution function from the mere knowledge of the one-particle distribution function. Note that this is true in spite of the fact that these particles do not \emph{interact}; because the random force is assumed to be prescribed, the time-evolution of $z_1$ is decoupled from the time-evolution of $z_2$.


\emph{Identifying the Langevin equation.}\,--- The one-particle Fokker-Planck equation associated with Eq.\,(\ref{ansatz}) is given by \cite{gardiner09,Bax84}
\begin{align}\label{fok_one_gen}
\frac{\partial f_t}{\partial t}&=-\text{div}(f_tX_0)+\frac{1}{2}\sum_{k\geq 1}\text{div}(\text{div}(f_t X_k)X_k)\nonumber\\
&=A_1f_t,                                     
\end{align}
while the two-particle Fokker-Planck equation \cite{Bax84,twopoint,kunita_lecture} is given by
\begin{align}\label{fok_two_gen}
\frac{\partial g_t}{\partial t}=&A_1^{(1)}g_t+A_1^{(2)}g_t\nonumber\\
                                           &+\sum_{k\geq 1}\text{div}^{(1)}\text{div}^{(2)}:g_tX_k(z_1)\otimes X_k(z_2).
\end{align}
The divergence operators in these expressions are defined relative to the Liouville volume form and the colon indicates the full contraction of second-rank tensors, $a:b\equiv a^{ij}b_{ij}$. Because these equations follow from Eq.\,(\ref{ansatz}) \emph{via} rigorous mathematics, we will refer to them as the mathematical Fokker-Planck equations.

On the other hand, under our assumption that the correlation time of the perturbed force is much shorter than a bounce time, standard coarse-graining procedures \cite{risken,diffusion_course} together with a decomposition theorem for time-ordered exponentials \cite{time_ordered} lead to the late-time evolution laws for the one- and two-particle distribution functions associated with the microscopic equations of motion, Eq.\,(\ref{micro}). The physical one-particle Fokker-Planck equation is given by 
\begin{align}\label{one_particle_fok}
\frac{\partial f_t}{\partial t}&=-\left\{f_t,H_0+\frac{\epsilon^2}{\tau}\mathbb{E}[s_2]\right\}+\frac{\epsilon^2}{2\tau}\mathbb{E}[\left\{\left\{f_t,s_1\right\},s_1\right\}]\nonumber\\
&=A_1f_t,                     
\end{align}  
while the physical two-particle Fokker-Planck equation (see the supplementary material for a derivation) is given by
\begin{align}\label{two_particle_fok}
\frac{\partial g_t}{\partial t}=A_1^{(1)}g_t+A_1^{(2)}g_t+\frac{\epsilon^2}{\tau}\mathbb{E}[\alpha:\mathbf{d}^{(1)}\mathbf{d}^{(2)}g_t].
\end{align}
The notation introduced in these two equations is defined as follows: $\mathbb{E}$ denotes an expectation value; the functions $s_1,s_2$ are defined by
\begin{subequations}
\label{kick}
\begin{align}
&s_1=\int_0^\tau\!\exp(\lambda X_{H_0})_{*}h_{\tau-\lambda}\mathrm{d}\lambda\label{s1}\\
&s_2=\frac{1}{2}\!\int_0^\tau\!\!\!\!\int_0^a\!\{\exp(bX_{H_0})_{*}h_{\tau-b},\exp(aX_{H_0})_{*}h_{\tau-a}\}\mathrm{d}b\,\mathrm{d}a;\label{s2}
\end{align}
\end{subequations}
$\exp(Y):M\rightarrow M$ denotes the time-one advance map of the dynamical system defined by the vector field $Y$; $(\exp(Y)_{*}h)(z)\equiv h(\exp(-Y)(z))$; the superscripts indicate which argument of $g_t$ that $A_1$ and the exterior derivative $\mathbf{d}$ should be applied to; and $\alpha(z_1,z_2)\equiv\mathbb{E}[X_{s_1}(z_1)\otimes X_{s_1}(z_2)]$ is the two-point covariance tensor. 

The $X_k$ must be chosen so that the mathematical Fokker-Planck equations, Eqs.\,(\ref{fok_one_gen}) and (\ref{fok_two_gen}), are equivalent to the physical Fokker-Planck equations, Eqs.\,(\ref{one_particle_fok}) and (\ref{two_particle_fok}). However, a direct comparison of these two pairs of equations is difficult with Eqs.\,(\ref{one_particle_fok}) and (\ref{two_particle_fok}) in their current form. To eliminate this issue, we will obtain a special decomposition of the two-point covariance tensor $\alpha(z_1,z_2)$. 

As a first step, notice that if we fix a one-form $\xi\in T_{z_1}^*M$, then we can define a vector field $Y_\xi$ on $M$ by contracting $\xi$ with $\alpha$ on the left according to 
\begin{align}\label{ydefined}
Y_\xi(z_2)&=\alpha(z_1,z_2)(\xi,\cdot)\nonumber\\
&=\mathbb{E}[\xi(X_{s_1}(z_1))X_{s_1}(z_2)].
\end{align}
By forming all possible linear combinations of vector fields of this form, we can construct a (potentially infinite dimensional) linear space of vector fields \cite{aro50,Bax76}, which we will denote $\mathcal{H}$,
\begin{align}\label{defH}
\mathcal{H}=\{\text{linear combinations of }Y_\xi,~\xi\in T^*M\}.
\end{align}
Because each $Y_\xi$ is of the form $Y_\xi(z)=X_{\bar{H}}(z)$ with $\bar{H}(z)=\mathbb{E}[\xi(X_{s_1}(z_o)){s_{1}}(z)]$, and the sum of Hamiltonian vector fields is again Hamiltonian, $\mathcal{H}$ consists entirely of Hamiltonian vector fields. Moreover, following Baxendale \cite{Bax84, Bax76}, we see that $\mathcal{H}$ is a real Hilbert space whose inner product is defined by the formula
\begin{align}\label{ip}
\left<Y_\xi,Y_\eta\right>_{\mathcal{H}}&=\alpha(z_1,z_2)(\xi,\eta)\nonumber\\
                                      &=\mathbb{E}[\xi(X_{s_1}(z_1))\eta(X_{s_1}(z_2))],
\end{align}
where $\xi\in T^*_{z_1}M$ and $\eta\in T^*_{z_2} M$. Therefore we may choose an orthonormal basis $\{e_k\}_{k\ge 1}$ for $\mathcal{H}$, where each $e_k$ must be of the form $e_k=X_{H_k}$. A simple calculation then leads to the desired decomposition of $\alpha$:
\begin{align}\label{decomp}
\alpha(z_1,z_2)=\sum_{k\ge 1}X_{H_k}(z_1)\otimes X_{H_k}(z_2).
\end{align}

Using this decomposition of the two-point covariance tensor, it is straightforward to manipulate Eqs.\,(\ref{one_particle_fok}) and (\ref{two_particle_fok}) into the same form as Eqs.\,(\ref{fok_one_gen}) and (\ref{fok_two_gen}). After doing so, it is trivial to identify the correct $X_k$. Indeed, we have found that the physical Langevin equation is given by
\begin{align}\label{result}
\delta z_t=X_{\tilde{H}_0}(z_t)\,\mathrm{d}t+\sum_{k\geq 1}X_{\tilde{H}_k}(z_t)\,\delta W^k_t,
\end{align}
where
\begin{align}
\tilde{H}_0=H_0+\frac{\epsilon^2}{\tau}\mathbb{E}[s_2],~~~\tilde{H}_k&=\frac{\epsilon}{\sqrt{\tau}}H_k\label{hks}
\end{align}
Recall that the $X_{H_k}$ are defined to be an orthonormal basis of the Hilbert space $\mathcal{H}$ defined in Eq.\,(\ref{defH}). Also recall that all of the above manipulations have been performed under the assumption that the correlation time of the perturbed force felt by a particle is much shorter than any bounce time associated with the perturbation.

Because the coefficients in the Langevin equation for stochastic acceleration, Eq.\,(\ref{result}), are all Hamiltonian vector fields, this equation is an example of a \emph{stochastic Hamiltonian system}, the foundations of which are developed in \cite{La08}. It is in this sense that the Langevin equation for stochastic acceleration inherits the Hamiltonian structure of the microscopic equations. In particular, SDEs of this type are known to arise from a stochastic variational principle for which Noether's theorem applies. Thus, even at the dissipative macroscopic level, symmetries imply the presence of conservation laws.


\emph{Example 1.}\,--- We will find the physical Langevin equation for two example stochastic acceleration problems. Generally speaking, finding the coefficients of the physical Langevin equation involves finding an orthonormal basis for the space $\mathcal{H}$, a task which may be analytically intractable. But, by Mercer's theorem \cite{mercer}, this task can be cast as an eigenvalue problem for which there are existing numerical solution methods. In any case, in these examples, the analytical route is tractable. 

First, consider a single-species, unmagnetized plasma subjected to a random weak electrostatic pulse at $\tau$-second intervals. Assume that the pulses are uniform in space and constant in magnitude, but uniformly and independently distributed in direction. Thus, the $k$'th pulse is generated by a potential of the form $\phi_k(\bm{x},t)=(\bm{z}_k\cdot \bm{x})\phi_o u(t-k\tau)$, where $\bm{z}_k$ is a random vector uniformly distributed over the unit sphere and $u(t)$ is a temporal windowing function localized at $t=\tau/2$.

In order to find the Langevin equation governing the plasma dynamics at times much longer than $\tau$, we must (a) calculate $s_1$ and $s_2$ using Eqs.\,(\ref{s1}) and (\ref{s2}), (b) find an orthonormal basis $\{X_{H_k}\}_{k\geq1}$ for the space $\mathcal{H}$ defined in Eq.\,(\ref{defH}), and (c) write down Eq.\,(\ref{result}) with $\tilde{H}_0$ and $\tilde{H}_k$ calculated using Eq.\,(\ref{hks}). The results of these three steps are as follows.

(a) A quick calculation shows that 
\begin{subequations}
\begin{align}
s_1&=m_o\bm{z}\cdot \bm{x}-m_1\bm{z}\cdot \bm{v}\\ 
s_2&=\text{const}
\end{align}
\end{subequations}
where $m_o=(q/m)\phi_o\int_0^\tau u(s)\mathrm{d}s$, $m_1=(q/m)\phi_o\int_0^\tau(\tau-s)u(s)\mathrm{d}s$, and $q/m$ is the charge-to-mass ratio.

(b) Each $Y_{\xi}$ must be of the form $Y_\xi=X_{g_{\bm{\beta}\bm{\gamma}}}$, where
\begin{align}
g_{\bm{\beta}\bm{\gamma}}(\bm{x},\bm{v})=\frac{1}{3}(m_1\bm{\beta}+m_o\bm{\gamma})\cdot(m_1\bm{v}-m_o\bm{x}),
\end{align}
and $\bm{\beta},\bm{\gamma}$ are arbitrary constant 3-component vectors. Using this expression, it is simple to find an orthonormal basis for $\mathcal{H}$. One is given by $\{X_{\bar{H}_k}\}_{k=1..3}$, with
\begin{align}
H_i(\bm{x},\bm{v})=\frac{1}{\sqrt{3}}e_i\cdot(m_1\bm{v}-m_o\bm{x}),
\end{align}
where $\{e_i\}_{i=1..3}$ is the standard basis for $\mathbb{R}^3$.

(c) Finally, the physical Langevin equation is given by
\begin{subequations}
\label{phys_lang}
\begin{align}
\delta x^i=v^i\,\mathrm{d}t+\frac{1}{\sqrt{3\tau}}m_1\,\delta W^i\\
\delta v^i=\frac{1}{\sqrt{3\tau}}m_o\,\delta W^i,
\end{align}
\end{subequations}
where $i=1,2,3$.

As is readily verified, the one-particle Fokker-Planck equation for this SDE is given by
\begin{align}\label{fpe}
\frac{\partial f_t}{\partial t}+v\cdot\nabla f_t=\frac{1}{6\tau}&(m_1^2\nabla^2f_t+m_om_1\nabla\cdot\nabla_vf_t\nonumber\\
&+m_om_1\nabla_v\cdot\nabla f_t+m_o^2\nabla_v^2f_t).
\end{align}
On the other hand, given an arbitrary function $\phi(\bm{x},\bm{v})$, the  SDE 
\begin{subequations}
\label{counter}
\begin{align}
\delta x^i=&v^i\,dt+\frac{m_1}{\sqrt{3\tau}}\left(\cos(\phi) \,\delta W^{1,i}-\sin(\phi) \,\delta W^{2,i}\right)\\
\delta v^i=&\frac{m_o}{\sqrt{3\tau}}\left(\cos(\phi) \,\delta W^{1,i}-\sin(\phi) \,\delta W^{2,i}\right),
\end{align}
\end{subequations}
where the $W^{1,i},W^{2,j}$ are six independent ordinary Wiener processes, will also generate Eq.\,(\ref{fpe}). However, when $\phi$ is not constant, the two-particle Fokker-Planck equation generated by Eq.\,(\ref{counter}) will differ from the two-point Fokker-Planck equation generated by Eq.\,(\ref{phys_lang}). This can be verified using Eq.\,(\ref{two_particle_fok}). The procedure identified here selects $\phi=0$ as the physical choice. In particular, it shows that a Langevin equation with the correct one-particle Fokker-Planck equation may still incorrectly reproduce the two-particle distribution function.

The inadequacy of Eq.\,(\ref{counter}) can also be understood intuitively as follows. Chaotic motions of any two particles experiencing the electrostatic pulses are ``synchronized" since the pulses are independent of $\bm{x}$ and $\bm{v}$. The Langevin equation (\ref{counter}), on the other hand, desynchronizes particle trajectories by involving additional Wiener processes, in spite of giving the correct one-particle Fokker-Planck equation.


\emph{Example 2.}\,--- Next, consider a minority population of magnetized fast ions moving through a plane lower-hybrid wave that propagates perpendicular to the magnetic field. Assume the wave has a high harmonic number and a wavelength small compared to a typical ion gyroradius. Karney \cite{karney79} has shown that the dynamics of the perpendicular velocity of these ions are governed by a canonical time-dependent Hamiltonian system with Hamiltonian
\begin{align}
H_t=I-\epsilon \sin(\sqrt{2I}\sin\theta-\nu t),
\end{align}
where $I$ is the normalized magnetic moment, $t$ the time normalized by the gyroperiod, $\theta$ the gyrophase, $\nu$ the harmonic number, and $\epsilon$ the normalized wave amplitude. Moreover, when $\epsilon$ exceeds a threshold value, an ion's motion becomes chaotic. This chaotic motion comes as the result of the effective randomization of the wave phase felt by an ion after a gyroperiod. Thus, above the threshold for chaos, we can model the wave phase as being randomized every gyroperiod by a random variable $\eta$. That is, we can replace the exact \emph{chaotic} ion motion with a stochastic approximation; see Ref.\,\cite{chirikov_universal} for Chirikov's application of the same modeling approach to the standard map. This allows us to apply the formalism developed in this Letter to find the physical Langevin equation describing the stochastic particle trajectories at times much longer than the gyroperiod.

As in the previous example, the first step is to calculate $s_1$ and $s_2$. Set $\tau=2\pi$ and adopt the rough approximation
\begin{align}\label{approx}
\sum_{n=-\infty}^\infty\frac{J_n}{\nu-n}\exp(in\theta)\approx\frac{J_{n_o}}{\delta}\exp(in_o\theta),
\end{align}
where $\nu=n_o+\delta$, $|\delta|<\frac{1}{2}$, and $J_n=J_n(\sqrt{2I})$ denotes the Bessel function of the first kind \cite{anb}. This approximation amounts to selecting the most slowly varying term in the sum in Eq.\,(\ref{approx}). Then, upon directly evaluating the integrals in Eqs.\,(\ref{s1}) and (\ref{s2}), the resulting expressions for $s_1$ and $\mathbb{E}[s_2]$ are
\begin{subequations}
\begin{align}
s_1&= 2\pi\text{sinc}(\pi\delta)J_{n_o}\sin(n_o\theta+\eta)\label{s1_2}\\
\mathbb{E}[s_2]&=\frac{\pi}{2}\sum_{m=-\infty}^\infty\frac{J_{m+1}^2-J_{m-1}^2}{m-\nu}\nonumber\\
                             &~~~+\frac{\pi}{2}\text{sinc}(2\pi\delta)\frac{J_{n_o+1}^2-J_{n_o-1}^2}{\delta},
\end{align}
\end{subequations}
where $\eta$ is a random variable uniformly distributed over the interval $[0,2\pi]$ and $\text{sinc}(x)=\sin(x)/x$.

Next, the space $\mathcal{H}$ can be constructed using the above expression for $s_1$. In this case, $\mathcal{H}$ is two-dimensional and has a basis $\{X_{H_1},X_{H_2}\}$, where
\begin{subequations}
\begin{align}
H_1(I,\theta)&=\sqrt{2}\pi\text{sinc}(\pi\delta) J_{n_o}(\sqrt{2I})\cos(n_o\theta)\\
H_2(I,\theta)&=\sqrt{2}\pi\text{sinc}(\pi\delta) J_{n_o}(\sqrt{2I})\sin(n_o\theta).
\end{align}
\end{subequations}

Finally, the coefficients for the Langevin equation, Eq.\,(\ref{result}), can be derived using Eq.\,(\ref{hks}). The result is
\begin{subequations}
\label{karney_sde}
\begin{align}
\delta I=&\epsilon\sqrt{\pi}\text{sinc}(\pi\delta)n_oJ_{n_o}(\sqrt{2I})\nonumber\\
&\times\left(\sin(n_o\theta)\delta W^1-\cos(n_o\theta)\delta W^2\right)\\
\delta \theta=&\left(1+\frac{\epsilon^2}{2\pi}\frac{\partial}{\partial I}\mathbb{E}[s_2]\right)\mathrm{d}t\nonumber\\
&+\bigg(\epsilon\sqrt{\frac{\pi}{2I}}\text{sinc}(\pi\delta)J_{n_o}^\prime(\sqrt{2I})\nonumber\\
&~~\times\left(\cos(n_o\theta)\delta W^1+\sin(n_o\theta)\delta W^2\right)\bigg).
\end{align}
\end{subequations}
The diffusion of the magnetic moment $I$ predicted by Eq.\,(\ref{karney_sde}) has already been studied by Karney \cite{karney79}. However, Eq.\,(\ref{karney_sde}) extends and compliments Karney's results by predicting the appropriate diffusion in gyrophase, as well as the correct two-particle statistics.


\emph{Concluding remarks.}\, --- We have shown how to derive the physical Langevin equation for particle trajectories undergoing stochastic acceleration. This SDE correctly generates the correct one- and two-particle Fokker-Planck equations and inherits the Hamiltonian structure of the microscopic equations of motion. This inheritance is theoretically satisfying because it is a direct consequence of demanding consistency with the physical one- and two-particle Fokker-Planck equations. It also implies that symmetries of the macroscopic physical laws governing stochastic acceleration imply the presence of conservation laws. While this relationship is well known at the microscopic level, it is a pleasant surprise that it remains intact upon passing to dissipative macroscopic equations.

A Hamiltonian Langevin equation \cite{La08} is a Stratonovich SDE of the form given in Eq.\,(\ref{result}). If a loop of initial conditions for this SDE evolves under a given realization of the noise, then the action of that loop is constant in time. In addition, these equations arise from a stochastic action principle \cite{La08} for which Noether's theorem applies. Thus, by showing the physical Langevin equation is Hamiltonian, we have also identified potentially powerful tools for the analysis of stochastic acceleration. In particular, using the methods of Bou-Rabee \cite{bou09}, the stochastic action principle can be used to develop variational integrators for Eq.\,(\ref{result}). Because these integrators are known to possess superior long-term statistical fidelity \cite{bou10}, this approach may prove to be useful in Monte Carlo simulations of stochastic acceleration.
 

%
%

%

\begin{acknowledgments}
The authors would like to express their appreciation to I. Dodin, J. A. Krommes, J. Parker, and G. W. Hammett. This work was supported by DOE contracts DE-AC02-09CH11466 and DE-FG02-04ER41289.
\end{acknowledgments}






\providecommand{\noopsort}[1]{}\providecommand{\singleletter}[1]{#1}%
%


\end{document}